\documentclass[prl,preprint,superscriptaddress,showpacs,amsmath,amssymb]{revtex4}

\usepackage{graphicx}
\usepackage{dcolumn}
\usepackage{bm}

\begin{document}

\title{From coupled elementary units to the complexity of the glass transition}


\date{\today}

\author{Christian Rehwald}
\affiliation{\frenchspacing Westf\"alische Wilhelms Universit\"at M\"unster, Institut f\"ur physikalische Chemie, Corrensstr.\ 30, 48149 M\"unster, Germany}
\affiliation{\frenchspacing Center of Nonlinear Science CeNoS, Westf\"alische Wilhelms Universit\"at M\"unster, Germany}
\author{Oliver Rubner}
\affiliation{\frenchspacing Westf\"alische Wilhelms Universit\"at M\"unster, Institut f\"ur physikalische Chemie, Corrensstr.\ 30, 48149 M\"unster, Germany}
\affiliation{\frenchspacing Center of Nonlinear Science CeNoS, Westf\"alische Wilhelms Universit\"at M\"unster, Germany}
\author{Andreas Heuer}
\affiliation{\frenchspacing Westf\"alische Wilhelms Universit\"at M\"unster, Institut f\"ur physikalische Chemie, Corrensstr.\ 30, 48149 M\"unster, Germany}
\affiliation{\frenchspacing Center of Nonlinear Science CeNoS, Westf\"alische Wilhelms Universit\"at M\"unster, Germany}


\begin{abstract}
 Supercooled liquids display fascinating properties upon cooling such as the emergence of dynamic length scales. Different models strongly vary with respect to the choice of the elementary subsystems (CRR) as well as their mutual coupling. Here we show via computer simulations of a glass former that both ingredients can be identified via analysis of finite-size effects within the continuous-time random walk framework. The CRR already contain complete information about thermodynamics and diffusivity whereas the coupling determines structural relaxation and the emergence of dynamic length scales.
\end{abstract}






\maketitle

Understanding the complex dynamics of supercooled liquids down
to the glass transition as reflected, e.g., by the emergence
of dynamic length scales  \cite{Berthier:2006,Stein08}  is still a highly controversal problem \cite{Debenedetti01,Dyre:2006}.
Qualitatively speaking the system decomposes into coupled elementary units  (sometimes
denoted cooperatively rearranging regions (CRRs) \cite{Adam65}). Presently discussed models differ dramatically
with respect to the nature of this decomposition. For example in the mosaic approach
\cite{Kirkpatrick89,Xia01,Wolynes:2001b,Lubchenko:2006} the complexity is fully
embedded in the properties of the CRRs, denoted mosaics, whereas in the other extreme
of the kinetically constrained models (KCMs)  \cite{Fredrickson84,Garrahan02,Jung2004,Hedges09}
the complexity is exclusively is related to the coupling while the elementary units are just trivial two-state systems.

Here we show that it is possible via computer simulations of a
standard model glass former to obtain a clear-cut decomposition
and thus to learn about the nature of the CRRs as well as the
coupling between them. The key idea is to directly extract
coupling effects from studying the size-dependence under periodic
boundary conditions. The analysis is performed in the
continuous-time random walk (CTRW) framework which for the first
time is quantitatively extended to larger systems. Two conclusions
can be drawn. Firstly, our results explain the fact that the
structural relaxation time displays major finite size effects
whereas there are hardly any for the diffusion constant. Secondly,
we present a minimum model of the glass transition which reflects
the observed coupling effects. Albeit similar in spirit to the
KCMs many important differences are present.

We have simulated the binary mixture Lennard-Jones system (BMLJ) which can be regarded as a prototype glass-former \cite{Kob1995}.
Simulations have been performed in the NVT ensemble.
For the smaller system we have have used a slightly shorter
cutoff of $r_c=1.8$ \cite{Doliwa:2003b}.
In previous work it has been shown that a system as small as $N_{min} = 65$ particles
with periodic boundary conditions basically displays very similar thermodynamic as well as diffusive
properties as a macroscopic system in
the range of temperatures, accessible to computer simulations (in
particular between $T_c$ and $2 T_c$) \cite{Heuer_Doliwa1}.
Interestingly, for the same temperature range $N=40$ is too small
\cite{Heuer_Buechner}. Thus, a BMLJ system with $N \approx
N_{min}$ is close to a system reflecting the effects of just a single CRR
(more precisely: two CRRs \cite{Heuer_trap}).

Due to the microscopic resolution computer simulations are very well suited
to extract relevant information about the coupling effects.
One route which has been chosen by Biroli et al. is to immobilize
a large system except for a sphere of variable radius $r$ and to
study the dynamics in this sphere in dependence of $r$
\cite{Biroli08}. One faces the problem to extract from this highly non-equilibrium
situation the relevant equilibrium properties. In this work
we analyze the size-dependence of the BMLJ system. From the perspective of a CRR
the coupling effects are switched on when increasing the system size.
Interestingly, the structural relaxation time $\tau_\alpha$
\cite{Stariolo06,Sastry09} as well as the length scale of dynamic
heterogeneities display significant finite-size effects
\cite{Whitelam2004,Berthier_Jack}.

A key observable to characterize the relaxation is the incoherent
scattering function $S(k,t)$ which is defined as
\begin{equation}
S(k,t) = \langle \exp(i \vec{k} [\Delta \vec{r}(t)] \rangle
\end{equation}
where  $\Delta \vec{r}$ denotes the single-particle displacement
vector during time $t$. The contribution of structural relaxation to the decay of $S(k,t)$ can be
determined by referring to the inherent structure (IS) trajectory,
obtained after minimizing every configuration \cite{Schroder:210},
reflecting the minima of the potential energy landscape (PEL) \cite{Wales:2003}. The resulting
incoherent scattering function is denoted
$S_{IS}(k,t)$. Qualitatively, this procedure corresponds to a removal of the vibrational contributions.
Conceptually similar is the use of the
metabasin (MB) trajectory \cite{Debenedetti01,Heuer_Doliwa1,Heuer_review}. A
MB consists of an appropriately chosen set of nearby IS.
The k-dependent relaxation time, which reflects
the time a particle on average needs to move the distance $2\pi/k$, is
defined via $\tau(k) = \int_0^\infty dt\, S_{IS}(k,t)$.

\begin{figure}[tb]
\centering\includegraphics[width=0.9\columnwidth]{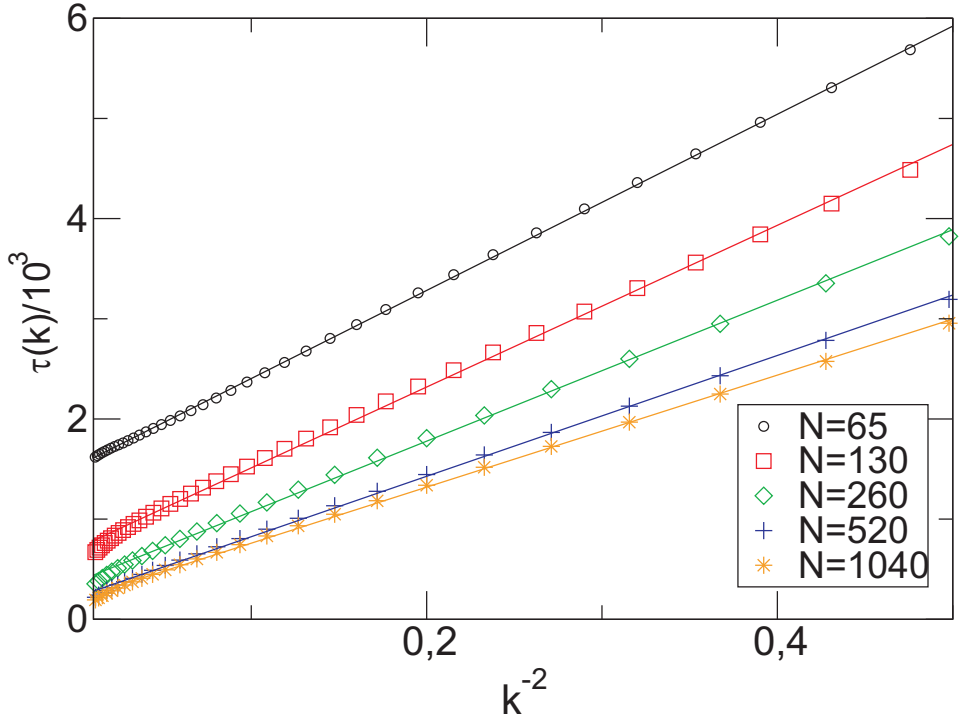}
\caption{The k-dependent relaxation time $\tau(k)$ vs $k^{-2}$ for
different system sizes at T=0.5. Included are linear fits which
work very well for $k \le 8$. From the two adjustable parameters
it is possible to derive the values of the diffusion constant $D$
and the structural relaxation time $\tau_\alpha$. } \label{fig1}
\end{figure}

In Fig.\ref{fig1} we show the wave vector dependent relaxation time $\tau(k)$
based on $S_{IS}(k,t)$ where the large k limit is significantly beyond the maximum of the structure factor.
As shown in \cite{Heuer_CTRW} the metabasin (MB) dynamics of a system with $N=N_{min}$ can be described
as a CTRW in configuration space (with waiting times $\tau_{MB}$) where spatial and temporal properties are decoupled
\cite{Heuer_CTRW}.  As a consequence
the k-dependence of $\tau(k)$ can be written  as \cite{Berthier:2005a,Heuer_CTRW}
\begin{equation}
\label{eqfit}
\tau(k)=\tau_\alpha+\frac{1}{3Dk^2}.
\end{equation}
where \cite{Berthier:2005a,Heuer_CTRW}
\begin{equation}
\label{eqmoment}
D \propto 1/\langle
\tau_{MB} \rangle, \, \, \tau_\alpha \propto D \langle \tau_{MB}^2
\rangle .
\end{equation}
Due to the presence of dynamic heterogeneities, the distribution of waiting times $\varphi(\tau_{MB})$ is
very broad so that $\tau_\alpha \gg \langle \tau_{MB} \rangle$.

\begin{figure}[tb]
\centering\includegraphics[width=0.9\columnwidth]{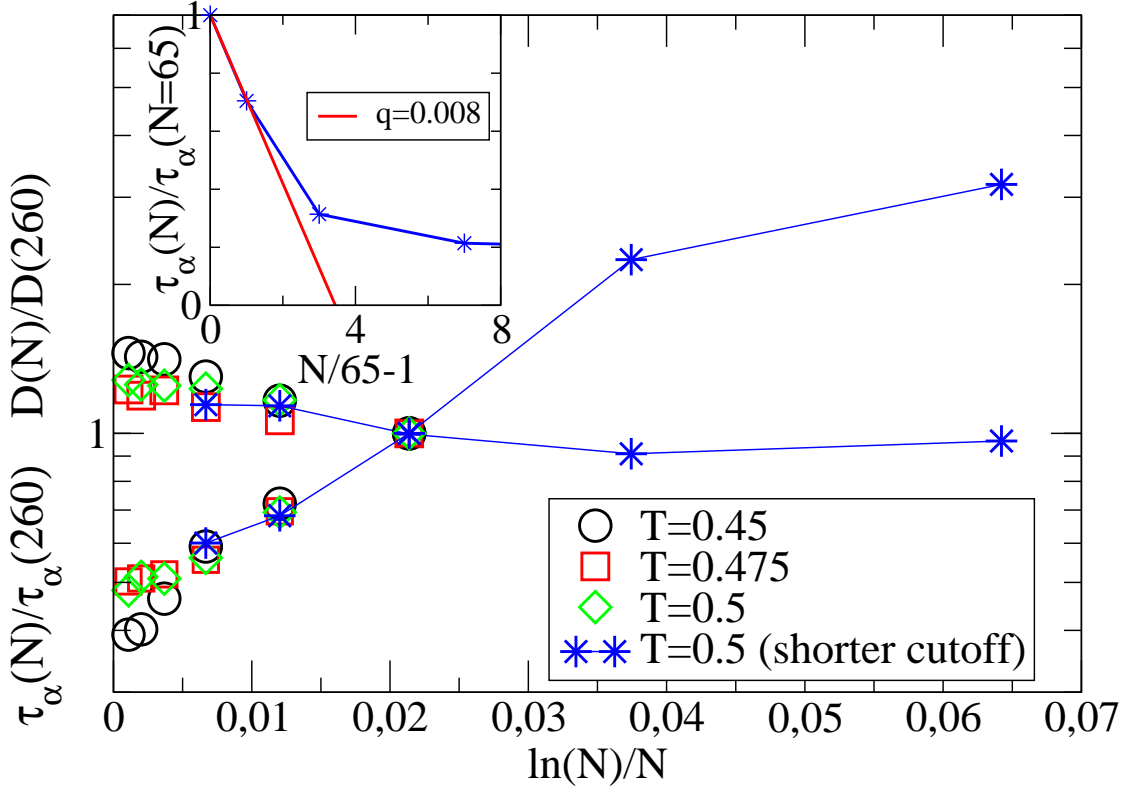}
\caption{The size-dependence of the diffusion constant $D$ and the
structural relaxation time $\tau_\alpha$ in relation to its value
for $N=260$. Whereas the diffusion constant only displays very
minor finite size effects, dramatic effects are observed for the
structural relaxation time. The inset contains a linear plot of
the finite-size effects for $\tau_\alpha$ together an estimation
of the coupling constant $q$.} \label{fig2}
\end{figure}

How to characterize the dynamics for large systems ($N \gg N_{min}$)? In this limit the MB
trajectory would violate one of the crucial assumptions of the
CTRW approach because due to the dynamic heterogeneities
successive MB transitions are often performed by identical
particles \cite{Heuer_CTRW}. In particular, Eq.\ref{eqmoment} is no longer valid for large $N$.
There exists, however, an elegant solution to this problem by defining local waiting times $\tau_{local}$
on the single-particle level in real space \cite{Vollmayr:2004,Hedges07}.
Specifically, one considers
a trajectory which is time-averaged
for a time somewhat longer than the vibrational time scale.
Whenever the particle has moved a fixed distance $d$ (here:
$d=1/3$ in LJ-units where the precise choice is irrelevant) one identifies a transition process.

Validity of the CTRW-behavior in configuration space implies that the corresponding single-particle
trajectories can be also described as a CTRW where
Eq.\ref{eqmoment} has to be modified by replacing $\tau_{MB}$ with $\tau_{local}$. One might expect that this CTRW-type single-particle dynamics should prevail also when increasing
the system size. Indeed, as can be seen from Fig.\ref{fig1} Eq.\ref{eqfit} is applicable for all $N$ in the complete $k$-range
under consideration.
In Fig.\ref{fig2} the resulting values of $D(N)$
and $\tau_\alpha(N)$ are shown for different temperatures. Note that the diffusion constant indeed
only displays small finite size effects whereas the relaxation time varies by nearly one order of magnitude.
At first glance this may seem somewhat counterintuitive because upon decreasing the system size major
finite-size effects are observed for a small-scale (large $k$) observable, i.e. $\tau_\alpha$, but not for
a large-scale (small $k$) observable, i.e. $D$.

\begin{figure}[tb]
\centering\includegraphics[width=0.9\columnwidth]{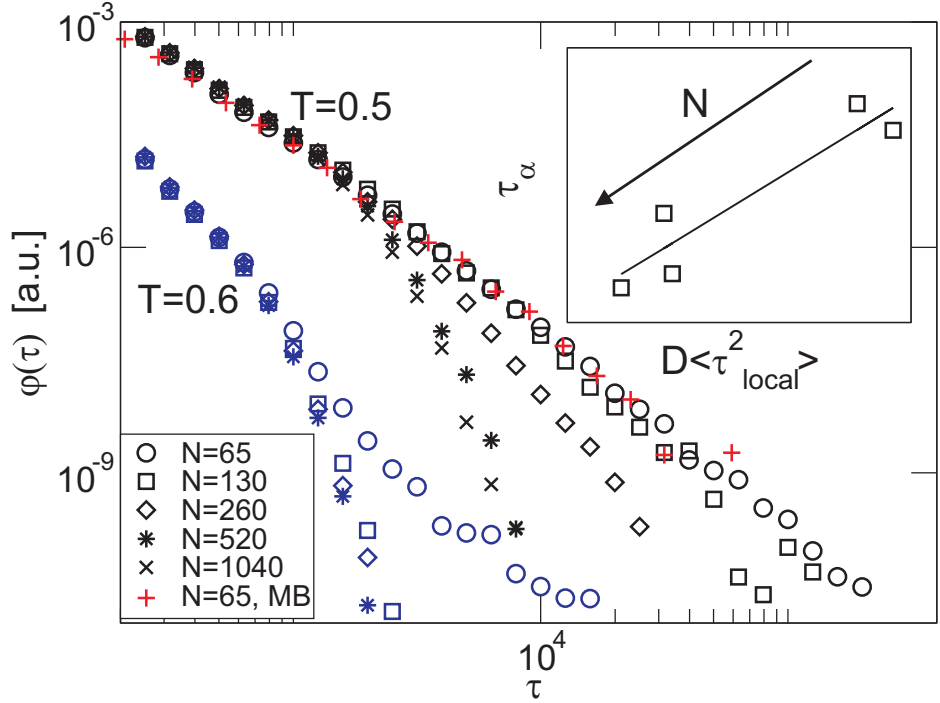}
\caption{The waiting time distribution
$\varphi(\tau_{local})$ for $T = 0.5$ and $T=0.6$ for different system sizes as well as $\varphi(\tau_{MB})$ for  $T=0.5$ and $N=65$. In the
inset the CTRW relation $\tau_\alpha \propto D \langle \tau^2_{local} \rangle$ is checked by comparing different system sizes between $N=65$ and $N=1040$ for $T=0.5$.}
\label{fig3}
\end{figure}

Finally, we analyse the $N$-dependence of
$\varphi(\tau_{local})$ as shown in Fig. \ref{fig3}. The data are discussed
in three different directions. First, for $N=N_{min}$ and $T=0.5$ we have
also included $\varphi(\tau_{MB})$ showing a very good agreement with $\varphi(\tau_{local})$,
thereby supporting the mapping of the waiting times in configuration space to those in real space.
The presence of a few extremely long
waiting times $(\tau_{local})$ reflects the fact that on average a
MB transition only covers half of the system size $N_{min}$ \cite{Heuer_trap}. Thus,
there is a finite probability that some particles remain immobile
even longer than the longest MB waiting time. Second, in the inset of
Fig.\ref{fig3} we verify the relation  in Eq.\ref{eqmoment} $\tau_\alpha \propto D
\langle \tau_{local}^2 \rangle$ within statistical uncertainty. Third, one clearly
sees that long residences of slow particles are strongly
suppressed when approaching the large-size limit. This observation
is particularly pronounced for the lower temperature where up to
$O(10^3)$ particles finite-size effects are present. One might be tempted to conclude that the
additional fluctuations around a given subsystem increase the
mobility and shifts all time scales to shorter times. This,
however, disagrees with the very small finite-size
effects of the diffusion constant which implies that the first
moment of the waiting time distribution is basically unchanged.

 For a system with
$N=N_{min}$ it has been shown that the state of the system can be,
to a reasonable approximation, characterized by a rate $\Gamma$,
denoting the probability for a MB transition
\cite{Heuer_trap,Heuer_review}. To discuss the nature of the
observed finite-size effects we assume for reasons of simplicity
that this approximation is strictly valid. Furthermore we
introduce $p(\Gamma)$ as the equilibrium distribution of rates. A
finite width of $p(\Gamma)$ is equivalent to the presence of
dynamic heterogeneities.  For $N=N_{min}$ it turns out that
$\Gamma$ is strongly related to the MB energy \cite{Heuer_review}. Thus, the observed
invariance of the thermodynamic properties, i.e. the trivial scaling of
$p(e)$ (e: MB energies) upon increasing the
system size, strongly suggests that $p(\Gamma)$ for the subsystem
does not change either.  This also follows from the previously
reported observation that in terms of MB transitions the total
system behaves as if it were a superposition of independent
subsystems of size $N_{min}$ \cite{doliwa:404}. As a consequence
the remaining effect of the coupling is to give rise to fluctuations
of the rate of the tagged subsystem due to relaxation processes in adjacent
subsystems without modifying the overall distribution $p(\Gamma)$
({\it passive exchange process}).

This coupling has one important consequence. During a very immobile period, characterized by a small
value of $\Gamma$, it is very likely that the system acquires a new
(and on average larger) value of $\Gamma$. This immediately
explains the numerical observation that the long-time contribution
to $\varphi(\tau_{local})$ disappears when the system size
increases and thus the fluctuations become faster.

Furthermore, these  observations allow us to understand the nature
of the finite-size effects of $S_{IS}(k,t)$. In particular one needs to
understand why finite-size effects occur for $\tau_\alpha$ rather than $D$.
Both observables are fundamentally different because (at least
for spatially uncorrelated jumps) the diffusivity is fully determined by the
properties of a single jump whereas for complete structural relaxation
basically every particle has to move, requiring a large number of
elementary relaxation processes. Thus, the presence of passive exchange
processes is relevant for $\tau_\alpha$ but not for $D$. A
more formal description of the resulting differences can be
achieved by using the relations  $D \propto \langle \Gamma
\rangle$ and $\tau_\alpha \propto \langle \Gamma^{-1} \rangle$
where the average is over $p(\Gamma)$ \cite{Diezemann:1998}. For
reasons of simplicity we consider the case where the fluctuations
of $\Gamma$ are extremely fast. Naturally, these fluctuations
leave the average rate $\langle \Gamma \rangle$ and thus $D$
unchanged. For $\langle \Gamma^{-1} \rangle$, however, $\Gamma$
needs to be substituted by $\langle \Gamma \rangle$. Then one
obtains
\begin{equation}
\tau_\alpha(N) = \langle \langle \Gamma \rangle ^{-1} \rangle =
1/\langle \Gamma \rangle < \langle \Gamma^{-1} \rangle =
\tau_\alpha(N=N_{min})
\end{equation}
where the inequality is a strict mathematical statement if dynamic
heterogeneities are present. Of course, the less pronounced the
fluctuations the smaller the ratio
$\tau_\alpha(N)/\tau_\alpha(N=N_{min})$.

\begin{figure}[tb]
\centering\includegraphics[width=0.9\columnwidth]{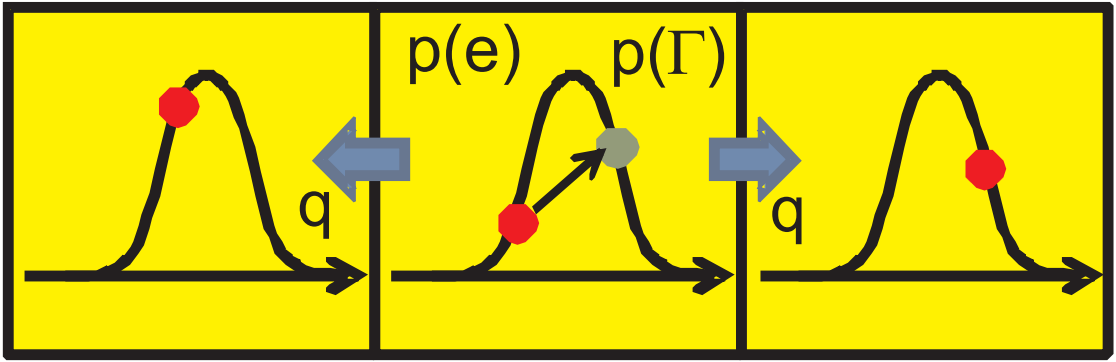} \caption{Schematic sketch of the physical scenario in supercooled liquids. After a MB transition
in the central subsystem the rates in the adjacent subsystems may change. In a minimum model with probably $q$ an adjacent subsystem
selects a randomly new rate according to the
Boltzmann probability and with probability $1-q$ there is no effect \cite{Heuer_review}.}
\label{fig4}
\end{figure}

A simple visualization of the physical scenario is presented in Fig.
\ref{fig4}:
(1) The local system of size $N_{min}$ fully contains the
information about the thermodynamics as well as the diffusion
constant (first moment of waiting time distribution). This system
can be very well characterized in the MB framework (given $p(e)$ and $\Gamma(e)$). Since, at
least in the analyzed temperature range, $N_{min}$ is roughly
temperature-independent no dynamic length scales are involved to
characterize, e.g., the temperature dependence of the diffusion
constant.(2) The macroscopic system can be regarded as a
superposition of systems of size $N_{min}$. The coupling is too
weak to change the thermodynamics (basically, this was the
criterion to define $N_{min}$) but gives rise to a dynamic
coupling between the subsystems. For the minimum model, sketched in Fig.
\ref{fig4} one can directly estimate the coupling constant $q$ in the limit
of small $q$ and not too large $N/N_{min}$ \cite{supporting} via the relation
\begin{equation}
\frac{\tau_\alpha(N)}{\tau_\alpha(N_{min})} = 1 - q \left ( \frac{N}{N_{min}}-1 \right ) \frac{\tau_\alpha(N_{min})}{\langle \tau \rangle}\left ( \frac{1}{\beta_M} - 1\right )
\end{equation}
with $\beta_M \approx 0.4$ and $\tau_\alpha(N=N_{min})/\langle \tau \rangle) \approx 30$ for the present system as determined
in Ref. \cite{Heuer_CTRW} in the CTRW context.
Comparison with the numerical data yields $q \approx 0.008$, see Fig.\ref{fig2}. This clearly shows that efficient coupling does not occur on the
time scale of MB transitions but for the present case on the time scale of $4\tau_\alpha(N=N_{min})$.

The resulting
correlations between the mobilities of adjacent systems will lead
to the emergence of temperature-dependent length scales of dynamic
heterogeneities. These correlations do not show up via a spatial correlation of instantaneous
rates but rather from the spatial correlation of the relaxation behavior during
sufficiently long time intervals (e.g. of the order of $\tau_\alpha$),
as exactly captured by the four-point correlation function \cite{Glotzer00}.
For a simple model
system one can show \cite{supporting} how the coupling modifies the value of
 $\tau_\alpha$ and how dynamic correlations and thus dynamic length scales emerge.

The aspect of coupling is rephrased in the KCMs in
a schematic way. Thus it is not surprising that
several similarities to atomic glass-formers exist (e.g. in the FA model the diffusion
constant does not depend on the coupling and the elementary length scale, i.e.
a spin, is temperature-independent \cite{Jung2004}). However, some important
differences remain. (1) Whereas a system
of size $N_{min}$, representing approx. two elementary units, already
reflects many features of the macroscopic system, the same number of spins
in the KCM naturally does not display any relevant
behavior. (2) In principle a subsystem can always relax without constraints
by its neighbors (see also Ref.\cite{Hedges09}) since the coupling is via passive exchange processes
rather than yes/no-decisions.  (3) Strong and fragile systems can
be understood by the same coupling rules by just changing the properties of the
potential energy landscape of the subsystems \cite{Heuer_review} whereas for the KCMs different rules have to be formulated.
(4) The experimentally observed correlation between thermodynamic and dynamic properties is kept whereas the thermodynamics
is considered irrelevant in the KCM approach.
(5) The coupling strength can be directly extracted from simulations of a atomistic glass-forming system, allowing a semi-quantitative
mapping between system and model.

In summary, the model, sketched above, can be, on the one hand, regarded as a generalization of the KCM whereas, on the
other hand, it is a minimum model, capturing a realistic version of the facilitation effect. In particular it is compatible
with the thermodynamic and dynamic observations for the model glass former reported in this work.
We just note that there is no obvious way to map the present results on the mosaic approach since the basic length scale, i.e. the
scale of the mosaics, is strongly temperature dependent.

We greatly acknowledge the financial support by the DFG (SFB 458).

\bibliography{doc}

\begin{thebibliography}{34}
\expandafter\ifx\csname natexlab\endcsname\relax\def\natexlab#1{#1}\fi
\expandafter\ifx\csname bibnamefont\endcsname\relax
  \def\bibnamefont#1{#1}\fi
\expandafter\ifx\csname bibfnamefont\endcsname\relax
  \def\bibfnamefont#1{#1}\fi
\expandafter\ifx\csname citenamefont\endcsname\relax
  \def\citenamefont#1{#1}\fi
\expandafter\ifx\csname url\endcsname\relax
  \def\url#1{\texttt{#1}}\fi
\expandafter\ifx\csname urlprefix\endcsname\relax\def\urlprefix{URL }\fi
\providecommand{\bibinfo}[2]{#2}
\providecommand{\eprint}[2][]{\url{#2}}

\bibitem[{\citenamefont{{{L. Berthier} {\it et. al}}}(2005)}]{Berthier:2006}
\bibinfo{author}{\bibnamefont{{{L. Berthier} {\it et. al}}}},
  \bibinfo{journal}{Science} \textbf{\bibinfo{volume}{310}},
  \bibinfo{pages}{1797} (\bibinfo{year}{2005}).

\bibitem[{\citenamefont{Stein and Andersen}(2008)}]{Stein08}
\bibinfo{author}{\bibfnamefont{R.~S.~L.} \bibnamefont{Stein}} \bibnamefont{and}
  \bibinfo{author}{\bibfnamefont{H.~C.} \bibnamefont{Andersen}},
  \bibinfo{journal}{Phys. Rev. Lett.} \textbf{\bibinfo{volume}{101}},
  \bibinfo{pages}{267802} (\bibinfo{year}{2008}).

\bibitem[{\citenamefont{{P. G. Debenedetti}{ and F. H.
  Stillinger}}(2001)}]{Debenedetti01}
\bibinfo{author}{\bibnamefont{{P. G. Debenedetti}{ and F. H. Stillinger}}},
  \bibinfo{journal}{Nature} \textbf{\bibinfo{volume}{410}}, \bibinfo{pages}{259
  } (\bibinfo{year}{2001}).

\bibitem[{\citenamefont{Dyre}(2006)}]{Dyre:2006}
\bibinfo{author}{\bibfnamefont{J.~C.} \bibnamefont{Dyre}},
  \bibinfo{journal}{Rev. Mod. Phys.} \textbf{\bibinfo{volume}{78}},
  \bibinfo{pages}{953} (\bibinfo{year}{2006}).

\bibitem[{\citenamefont{Adam and Gibbs}(1965)}]{Adam65}
\bibinfo{author}{\bibfnamefont{G.}~\bibnamefont{Adam}} \bibnamefont{and}
  \bibinfo{author}{\bibfnamefont{J.~H.} \bibnamefont{Gibbs}},
  \bibinfo{journal}{J. Chem. Phys.} \textbf{\bibinfo{volume}{43}},
  \bibinfo{pages}{139} (\bibinfo{year}{1965}).

\bibitem[{\citenamefont{Kirkpatrick et~al.}(1989)\citenamefont{Kirkpatrick,
  Thirumalai, and Wolynes}}]{Kirkpatrick89}
\bibinfo{author}{\bibfnamefont{T.~R.} \bibnamefont{Kirkpatrick}},
  \bibinfo{author}{\bibfnamefont{D.}~\bibnamefont{Thirumalai}},
  \bibnamefont{and} \bibinfo{author}{\bibfnamefont{P.~G.}
  \bibnamefont{Wolynes}}, \bibinfo{journal}{Phys. Rev. A}
  \textbf{\bibinfo{volume}{40}}, \bibinfo{pages}{1045} (\bibinfo{year}{1989}).

\bibitem[{\citenamefont{Xia and Wolynes}(2001{\natexlab{a}})}]{Xia01}
\bibinfo{author}{\bibfnamefont{X.}~\bibnamefont{Xia}} \bibnamefont{and}
  \bibinfo{author}{\bibfnamefont{P.~G.} \bibnamefont{Wolynes}},
  \bibinfo{journal}{Phys. Rev. Lett.} \textbf{\bibinfo{volume}{86}},
  \bibinfo{pages}{5526} (\bibinfo{year}{2001}{\natexlab{a}}).

\bibitem[{\citenamefont{Xia and Wolynes}(2001{\natexlab{b}})}]{Wolynes:2001b}
\bibinfo{author}{\bibfnamefont{X.}~\bibnamefont{Xia}} \bibnamefont{and}
  \bibinfo{author}{\bibfnamefont{P.~G.} \bibnamefont{Wolynes}},
  \bibinfo{journal}{J. Phys. Chem. B} \textbf{\bibinfo{volume}{105}},
  \bibinfo{pages}{6570} (\bibinfo{year}{2001}{\natexlab{b}}).

\bibitem[{\citenamefont{Lubchenko and Wolynes}(2006)}]{Lubchenko:2006}
\bibinfo{author}{\bibfnamefont{V.}~\bibnamefont{Lubchenko}} \bibnamefont{and}
  \bibinfo{author}{\bibfnamefont{P.~G.} \bibnamefont{Wolynes}},
  \bibinfo{journal}{Ann. Rev. Phys. Chem.} \textbf{\bibinfo{volume}{58}},
  \bibinfo{pages}{235} (\bibinfo{year}{2006}).

\bibitem[{\citenamefont{Fredrickson and Andersen}(1984)}]{Fredrickson84}
\bibinfo{author}{\bibfnamefont{G.~H.} \bibnamefont{Fredrickson}}
  \bibnamefont{and} \bibinfo{author}{\bibfnamefont{H.~C.}
  \bibnamefont{Andersen}}, \bibinfo{journal}{Phys. Rev. Lett.}
  \textbf{\bibinfo{volume}{53}}, \bibinfo{pages}{1244} (\bibinfo{year}{1984}).

\bibitem[{\citenamefont{Garrahan and Chandler}(2002)}]{Garrahan02}
\bibinfo{author}{\bibfnamefont{J.~P.} \bibnamefont{Garrahan}} \bibnamefont{and}
  \bibinfo{author}{\bibfnamefont{D.}~\bibnamefont{Chandler}},
  \bibinfo{journal}{Phys. Rev. Lett.} \textbf{\bibinfo{volume}{89}},
  \bibinfo{pages}{035704} (\bibinfo{year}{2002}).

\bibitem[{\citenamefont{Jung et~al.}(2004)\citenamefont{Jung, Garrahan, and
  Chandler}}]{Jung2004}
\bibinfo{author}{\bibfnamefont{Y.~J.} \bibnamefont{Jung}},
  \bibinfo{author}{\bibfnamefont{J.~P.} \bibnamefont{Garrahan}},
  \bibnamefont{and} \bibinfo{author}{\bibfnamefont{D.}~\bibnamefont{Chandler}},
  \bibinfo{journal}{Phys. Rev. E} \textbf{\bibinfo{volume}{69}},
  \bibinfo{pages}{061205} (\bibinfo{year}{2004}).

\bibitem[{\citenamefont{Hedges et~al.}(2009)\citenamefont{Hedges, Jack,
  Garrahan, and Chandler}}]{Hedges09}
\bibinfo{author}{\bibfnamefont{L.~O.} \bibnamefont{Hedges}},
  \bibinfo{author}{\bibfnamefont{R.~L.} \bibnamefont{Jack}},
  \bibinfo{author}{\bibfnamefont{J.~P.} \bibnamefont{Garrahan}},
  \bibnamefont{and} \bibinfo{author}{\bibfnamefont{D.}~\bibnamefont{Chandler}},
  \bibinfo{journal}{Science} \textbf{\bibinfo{volume}{323}},
  \bibinfo{pages}{1309} (\bibinfo{year}{2009}).

\bibitem[{\citenamefont{Kob and Andersen}(1995)}]{Kob1995}
\bibinfo{author}{\bibfnamefont{W.}~\bibnamefont{Kob}} \bibnamefont{and}
  \bibinfo{author}{\bibfnamefont{H.~C.} \bibnamefont{Andersen}},
  \bibinfo{journal}{Phys. Rev. E} \textbf{\bibinfo{volume}{52}},
  \bibinfo{pages}{4134} (\bibinfo{year}{1995}).

\bibitem[{\citenamefont{Doliwa and Heuer}(2003{\natexlab{a}})}]{Doliwa:2003b}
\bibinfo{author}{\bibfnamefont{B.}~\bibnamefont{Doliwa}} \bibnamefont{and}
  \bibinfo{author}{\bibfnamefont{A.}~\bibnamefont{Heuer}},
  \bibinfo{journal}{Phys. Rev. E} \textbf{\bibinfo{volume}{67}},
  \bibinfo{pages}{030501} (\bibinfo{year}{2003}{\natexlab{a}}).

\bibitem[{\citenamefont{Doliwa and Heuer}(2003{\natexlab{b}})}]{Heuer_Doliwa1}
\bibinfo{author}{\bibfnamefont{B.}~\bibnamefont{Doliwa}} \bibnamefont{and}
  \bibinfo{author}{\bibfnamefont{A.}~\bibnamefont{Heuer}},
  \bibinfo{journal}{Phys. Rev. E} \textbf{\bibinfo{volume}{67}},
  \bibinfo{pages}{031506} (\bibinfo{year}{2003}{\natexlab{b}}).

\bibitem[{\citenamefont{Buechner and Heuer}(1999)}]{Heuer_Buechner}
\bibinfo{author}{\bibfnamefont{S.}~\bibnamefont{Buechner}} \bibnamefont{and}
  \bibinfo{author}{\bibfnamefont{A.}~\bibnamefont{Heuer}},
  \bibinfo{journal}{Phys. Rev. E} \textbf{\bibinfo{volume}{60}},
  \bibinfo{pages}{6507} (\bibinfo{year}{1999}).

\bibitem[{\citenamefont{Heuer et~al.}(2005)\citenamefont{Heuer, Doliwa, and
  Saksaengwijit}}]{Heuer_trap}
\bibinfo{author}{\bibfnamefont{A.}~\bibnamefont{Heuer}},
  \bibinfo{author}{\bibfnamefont{B.}~\bibnamefont{Doliwa}}, \bibnamefont{and}
  \bibinfo{author}{\bibfnamefont{A.}~\bibnamefont{Saksaengwijit}},
  \bibinfo{journal}{Phys. Rev. E} \textbf{\bibinfo{volume}{72}},
  \bibinfo{pages}{021503} (\bibinfo{year}{2005}).

\bibitem[{\citenamefont{Biroli et~al.}(2008)\citenamefont{Biroli, Bouchaud,
  Cavagna, Grigera, and Verrocchio}}]{Biroli08}
\bibinfo{author}{\bibfnamefont{G.}~\bibnamefont{Biroli}},
  \bibinfo{author}{\bibfnamefont{J.~P.} \bibnamefont{Bouchaud}},
  \bibinfo{author}{\bibfnamefont{A.}~\bibnamefont{Cavagna}},
  \bibinfo{author}{\bibfnamefont{T.~S.} \bibnamefont{Grigera}},
  \bibnamefont{and}
  \bibinfo{author}{\bibfnamefont{P.}~\bibnamefont{Verrocchio}},
  \bibinfo{journal}{Nature Physics} \textbf{\bibinfo{volume}{4}},
  \bibinfo{pages}{771} (\bibinfo{year}{2008}).

\bibitem[{\citenamefont{Stariolo and Fabricius}(2006)}]{Stariolo06}
\bibinfo{author}{\bibfnamefont{D.~A.} \bibnamefont{Stariolo}} \bibnamefont{and}
  \bibinfo{author}{\bibfnamefont{G.}~\bibnamefont{Fabricius}},
  \bibinfo{journal}{J. Chem. Phys.} \textbf{\bibinfo{volume}{125}},
  \bibinfo{pages}{064505} (\bibinfo{year}{2006}).

\bibitem[{\citenamefont{Karmakar et~al.}(2009)\citenamefont{Karmakar, Dasgupta,
  and Sastry}}]{Sastry09}
\bibinfo{author}{\bibfnamefont{S.}~\bibnamefont{Karmakar}},
  \bibinfo{author}{\bibfnamefont{C.}~\bibnamefont{Dasgupta}}, \bibnamefont{and}
  \bibinfo{author}{\bibfnamefont{S.}~\bibnamefont{Sastry}},
  \bibinfo{journal}{Proc. Natl. Acad. Sci.} \textbf{\bibinfo{volume}{106}},
  \bibinfo{pages}{3675} (\bibinfo{year}{2009}).

\bibitem[{\citenamefont{Whitelam et~al.}(2004)\citenamefont{Whitelam, Berthier,
  and Garrahan}}]{Whitelam2004}
\bibinfo{author}{\bibfnamefont{S.~P.} \bibnamefont{Whitelam}},
  \bibinfo{author}{\bibfnamefont{L.}~\bibnamefont{Berthier}}, \bibnamefont{and}
  \bibinfo{author}{\bibfnamefont{J.~P.} \bibnamefont{Garrahan}},
  \bibinfo{journal}{Phys. Rev. Lett.} \textbf{\bibinfo{volume}{92}},
  \bibinfo{pages}{185705} (\bibinfo{year}{2004}).

\bibitem[{\citenamefont{Berthier and Jack}(2007)}]{Berthier_Jack}
\bibinfo{author}{\bibfnamefont{L.}~\bibnamefont{Berthier}} \bibnamefont{and}
  \bibinfo{author}{\bibfnamefont{R.~L.} \bibnamefont{Jack}},
  \bibinfo{journal}{Phys. Rev. E} \textbf{\bibinfo{volume}{76}},
  \bibinfo{pages}{041509} (\bibinfo{year}{2007}).

\bibitem[{\citenamefont{Schr{\o}der et~al.}(2000)\citenamefont{Schr{\o}der,
  Sastry, Dyre, and Glotzer}}]{Schroder:210}
\bibinfo{author}{\bibfnamefont{T.~B.} \bibnamefont{Schr{\o}der}},
  \bibinfo{author}{\bibfnamefont{S.}~\bibnamefont{Sastry}},
  \bibinfo{author}{\bibfnamefont{J.~C.} \bibnamefont{Dyre}}, \bibnamefont{and}
  \bibinfo{author}{\bibfnamefont{S.~C.} \bibnamefont{Glotzer}},
  \bibinfo{journal}{J. Chem. Phys.} \textbf{\bibinfo{volume}{112}},
  \bibinfo{pages}{9834} (\bibinfo{year}{2000}).

\bibitem[{\citenamefont{Wales}(2003)}]{Wales:2003}
\bibinfo{author}{\bibfnamefont{D.~J.} \bibnamefont{Wales}},
  \emph{\bibinfo{title}{Energy landscapes}} (\bibinfo{publisher}{Cambridge
  University Press}, \bibinfo{year}{2003}).

\bibitem[{\citenamefont{Heuer}(2008)}]{Heuer_review}
\bibinfo{author}{\bibfnamefont{A.}~\bibnamefont{Heuer}}, \bibinfo{journal}{J.
  Phys.: Cond. Mat.} \textbf{\bibinfo{volume}{20}}, \bibinfo{pages}{373101}
  (\bibinfo{year}{2008}).

\bibitem[{\citenamefont{Rubner and Heuer}(2008)}]{Heuer_CTRW}
\bibinfo{author}{\bibfnamefont{O.}~\bibnamefont{Rubner}} \bibnamefont{and}
  \bibinfo{author}{\bibfnamefont{A.}~\bibnamefont{Heuer}},
  \bibinfo{journal}{Phys. Rev. E} \textbf{\bibinfo{volume}{78}},
  \bibinfo{pages}{011504} (\bibinfo{year}{2008}).

\bibitem[{\citenamefont{Berthier et~al.}(2005)\citenamefont{Berthier, Chandler,
  and Garrahan}}]{Berthier:2005a}
\bibinfo{author}{\bibfnamefont{L.}~\bibnamefont{Berthier}},
  \bibinfo{author}{\bibfnamefont{D.}~\bibnamefont{Chandler}}, \bibnamefont{and}
  \bibinfo{author}{\bibfnamefont{J.~P.} \bibnamefont{Garrahan}},
  \bibinfo{journal}{Europhys. Lett.} \textbf{\bibinfo{volume}{69}},
  \bibinfo{pages}{320} (\bibinfo{year}{2005}).

\bibitem[{\citenamefont{Vollmayr-Lee}(2004)}]{Vollmayr:2004}
\bibinfo{author}{\bibfnamefont{K.}~\bibnamefont{Vollmayr-Lee}},
  \bibinfo{journal}{J. Chem. Phys} \textbf{\bibinfo{volume}{121}},
  \bibinfo{pages}{4781} (\bibinfo{year}{2004}).

\bibitem[{\citenamefont{Hedges et~al.}(2007)\citenamefont{Hedges, Maibaum,
  Chandler, and Garrahan}}]{Hedges07}
\bibinfo{author}{\bibfnamefont{L.~O.} \bibnamefont{Hedges}},
  \bibinfo{author}{\bibfnamefont{L.}~\bibnamefont{Maibaum}},
  \bibinfo{author}{\bibfnamefont{D.}~\bibnamefont{Chandler}}, \bibnamefont{and}
  \bibinfo{author}{\bibfnamefont{J.~P.} \bibnamefont{Garrahan}},
  \bibinfo{journal}{J. Chem. Phys.} \textbf{\bibinfo{volume}{237}},
  \bibinfo{pages}{211101} (\bibinfo{year}{2007}).

\bibitem[{\citenamefont{Doliwa and Heuer}(2003{\natexlab{c}})}]{doliwa:404}
\bibinfo{author}{\bibfnamefont{B.}~\bibnamefont{Doliwa}} \bibnamefont{and}
  \bibinfo{author}{\bibfnamefont{A.}~\bibnamefont{Heuer}}, \bibinfo{journal}{J.
  Phys. C: Cond. Mat.} \textbf{\bibinfo{volume}{15}}, \bibinfo{pages}{S849}
  (\bibinfo{year}{2003}{\natexlab{c}}).

\bibitem[{\citenamefont{Diezemann et~al.}(1998)\citenamefont{Diezemann,
  Sillescu, Hinze, and B{\"o}hmer}}]{Diezemann:1998}
\bibinfo{author}{\bibfnamefont{G.}~\bibnamefont{Diezemann}},
  \bibinfo{author}{\bibfnamefont{H.}~\bibnamefont{Sillescu}},
  \bibinfo{author}{\bibfnamefont{G.}~\bibnamefont{Hinze}}, \bibnamefont{and}
  \bibinfo{author}{\bibfnamefont{R.}~\bibnamefont{B{\"o}hmer}},
  \bibinfo{journal}{Phys. Rev. E} \textbf{\bibinfo{volume}{57}},
  \bibinfo{pages}{4398} (\bibinfo{year}{1998}).

\bibitem[{sup()}]{supporting}
\bibinfo{howpublished}{See EPAPS Document No.}

\bibitem[{\citenamefont{Glotzer et~al.}(2000)\citenamefont{Glotzer, Novikov,
  and Schroder}}]{Glotzer00}
\bibinfo{author}{\bibfnamefont{S.~C.} \bibnamefont{Glotzer}},
  \bibinfo{author}{\bibfnamefont{V.~N.} \bibnamefont{Novikov}},
  \bibnamefont{and} \bibinfo{author}{\bibfnamefont{T.~B.}
  \bibnamefont{Schroder}}, \bibinfo{journal}{J. Chem. Phys.}
  \textbf{\bibinfo{volume}{112}}, \bibinfo{pages}{509} (\bibinfo{year}{2000}).

\end{thebibliography}

\newpage

{\bf \large Supporting Online Material}

\vspace{1cm}

{\bf \it Discussion of a simple but illuminating model}

\begin{figure}[tb]
\centering\includegraphics[width=1.0\columnwidth]{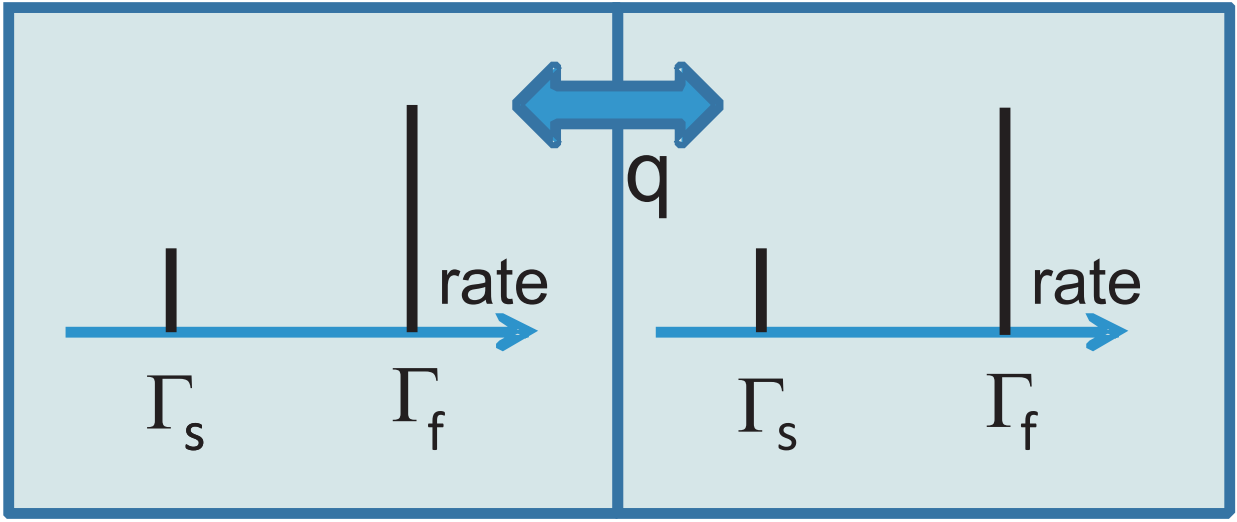} \caption{(Fig.S1) A model system with two subsystems; see text for details.}
\label{graph0}
\end{figure}

To substantiate the findings, related to the coupling effect, we
analyze a minimum model which leaves the thermodynamics  as well
as the diffusivity unchanged upon coupling.  It consists of two subsystems
with a coupling constant $q$, characterizing the strength of the coupling.
The model is sketched in Fig.S1. The elementary subsystem contains two
states. The slow escape rate from the lower state is $\Gamma_s$,
the faster escape rate from the upper state is $\Gamma_f$. The
relevance of  dynamical heterogeneities is characterized by
$\Gamma_f/\Gamma_s$. The degeneracy of the high-energy state is
proportional to $\varphi_f$, that of the lower state to
$\varphi_s$ ($\varphi_s + \varphi_f = 1$).  The Boltzmann
probabilities $p_i$ are proportional to $\varphi_i/\Gamma_i$. For
reasons of simplicity we choose $p_f = p_s = 1/2$ which implies
$\varphi_i= \Gamma_i/(\Gamma_s + \Gamma_f)$. For this subsystem
one directly obtains $\langle \tau \rangle =  2/(\Gamma_s +
\Gamma_f)$ and $\tau_\alpha = \langle \tau^2 \rangle / (2 \langle
\tau \rangle) = (1/2)(1/\Gamma_s + 1/\Gamma_f$). Now we consider a
larger system, containing two of these subsystems which are dynamically
coupled.  A relaxation
of, e.g., subsystem 1 has two consequences. First, subsystem 1
acquires a randomly new chosen state with probability $\varphi_i$.
This condition just reflects the independence of two subsequent
states, populated by subsystem 1. Second, with probability $q$
subsystem 2 randomly chooses a  state with probability $p_i$
whereas with probability $1-q$ no change occurs. The value of $q$
determines the strength of the coupling ($q=0$: no coupling, $q=1$
maximum coupling). This procedure guarantees that the Boltzmann
distribution in one subsystem and the average waiting time
does not change as a consequence of
the coupling to dynamical processes in the other subsystem.

In what follows we show via straightforward Monte Carlo simulations in
combination with analytical expressions that (i)
$\tau_\alpha$ indeed decreases with increasing system size and
that (ii) the coupling gives rise to dynamic length scales.
The key observable, recorded during the Monte Carlo simulations
are the different moments $\langle \tau^n
\rangle$ of the waiting time distribution of a single subsystem.
$\tau_\alpha$ is defined as $\int \, dt S_0(t) = \langle \tau^2
\rangle /(2 \langle \tau \rangle)$ where $S_0(t)$ denotes the
probability that a subsystem does not perform a relaxation
process during a time interval of length $t$.

\begin{figure}[tb]
\centering\includegraphics[width=1.0\columnwidth]{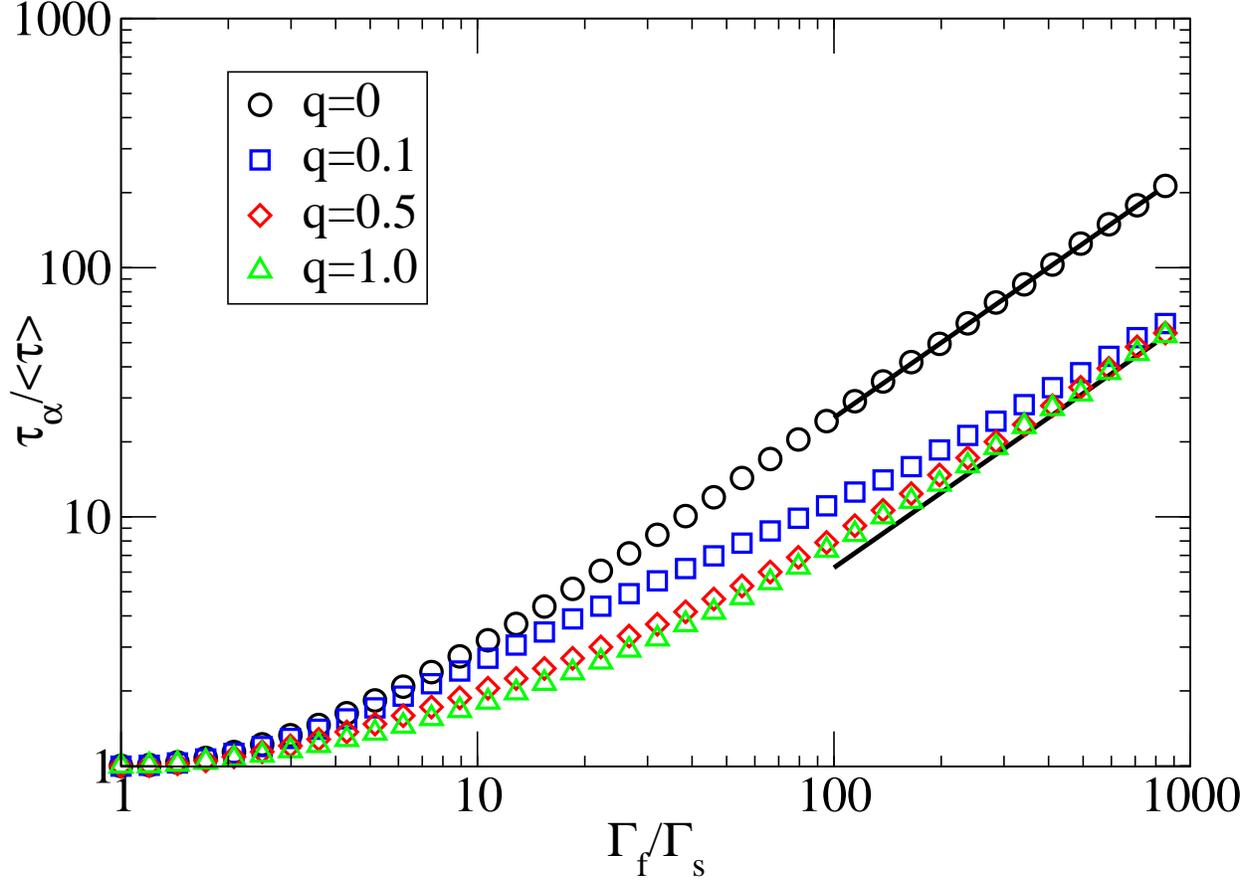} \caption{(Fig.S2)The dependence of $\tau_\alpha/\langle \tau \rangle$ on $\Gamma_f/\Gamma_s$.
Included are theoretical predictions $\tau_\alpha/\langle \tau \rangle = \Gamma_f/(4\Gamma_s)$ and $\tau_\alpha/\langle \tau \rangle = \Gamma_f/(16\Gamma_s)$ for the limit of large dynamic heterogeneities.}
\label{graph1}
\end{figure}

In Fig.S2 the
dependence of $\tau_\alpha/\langle \tau \rangle$ is shown as a
function of $\Gamma_f/\Gamma_s$ for different values of $q$. We
checked that in agreement with the setup of the model the value of
$\langle \tau \rangle \propto 1/D$ does not depend on $q$. Several
interesting observations can be made. (1) $\tau_\alpha$ decreases
with increasing coupling. This naturally reflects the general
system size dependencies of $\tau_\alpha$, reported in the main
text. (2) For large dynamic heterogeneities the q-dependence
disappears as long as $q > 0$, indicating some degree of
universality for large dynamic heterogeneities. (3) In this limit
one obtains $\tau_\alpha/\langle \tau \rangle =
\Gamma_f/(4\Gamma_s)$ for $q=0$ and $\tau_\alpha/\langle \tau
\rangle = \Gamma_f/(16\Gamma_s)$ for $q>0$. These relations can be
easily rationalized. For $q=0$ and $\Gamma_f \gg \Gamma_s$ the
relevant long-time decay of $S_0(t)$ is given by $(1/2)
\exp(-\Gamma_s t)$, implying $\tau_\alpha \approx 1/(2 \Gamma_s)$.
Together with $\langle \tau \rangle \approx 2/\Gamma_f$ the
observed relation immediately follows. For $q > 0$  a slow
relaxation process is only observed if for $t=0$ {\it both}
subsystems are in the slow state. If we define $n_{12}(t)$ as the
probability that none of the systems has performed a relaxation process
until time $t$ one naturally has  $n_{12}(t) = (1/4) \exp(-2\Gamma_s t)$.
The prefactor expresses the fact that only in one out of four cases both subsystems
are initially slow.
After the relaxation of, e.g., subsystem 1 and as a consequence of  $\varphi_f \gg \varphi_s$
subsystem 1 will typically
become fast afterwards. The large number of subsequent relaxation
processes will for $q > 0$ finally give rise to an exchange of the
slow to the fast state in subsystem 2. Thus, for $q\Gamma_f \gg \Gamma_s$ the time
difference between the initial relaxation process in subsystem 1 and the
first relaxation process in subsystem 2 is much smaller than $1/\Gamma_s$.
As a consequence one has $S_0(t) \approx n_{12}(t)$, yielding
 $\tau_\alpha/\langle \tau \rangle = \Gamma_f/(16\Gamma_s)$ as
indeed observed.

\begin{figure}[tb]
\centering\includegraphics[width=1.0\columnwidth]{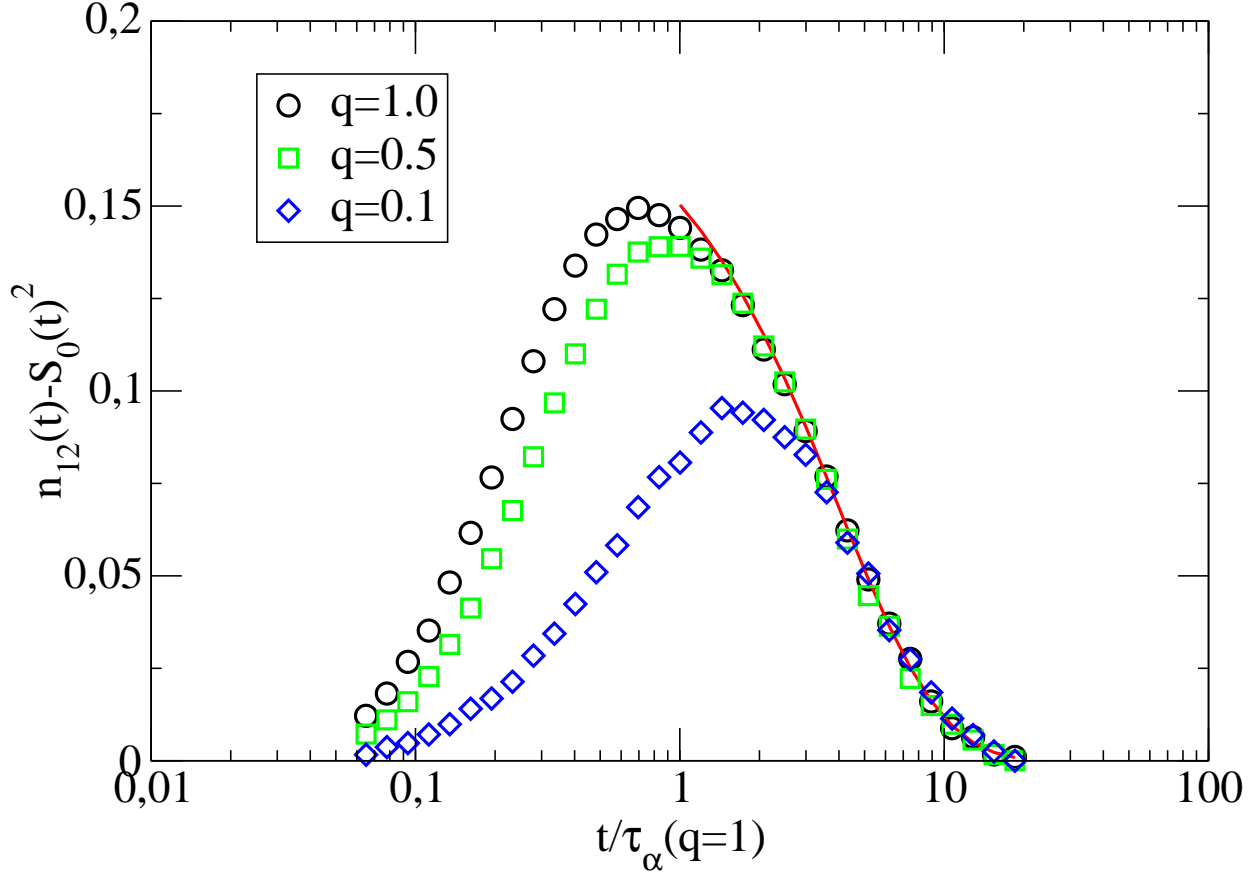} \caption{(Fig.S3)In analogy to
the four-point correlation functions the dynamic
correlations for the present system can be characterized  via $ n_{12}(t)  - [S_0(t)]^2$ where $
n_{12}(t) $ denotes the probability that the total system
does not relax during a time interval of length $t$. The solid lines correspond to the theoretical
expectation for long times. }
\label{graph2}
\end{figure}

At a particular time the rates in both subsystems are
uncorrelated to each other. However, for $q > 0$ both subsystems are dynamically
correlated if observed for a finite time interval; see Fig.S3. In agreement with typical data for glass-forming
systems one observes a maximum correlation for $t \approx
\tau_\alpha$. In analogy to above one expects for $t >
\tau_\alpha$ that $S_0(t) = n_{12}(t)= (1/4) \exp(-2 \Gamma_s t)$
which agrees with the numerical data.

\vspace{1cm}

{\bf \it Finite-size effects for small coupling constant $q$}

In generalization to above we consider two subsystems with a general distribution of rates $p(\Gamma)$.
For one tagged system we write for the relaxation function (in discrete notation) $S(t) = \sum_i p_i g_i(t)$
where $p_i$ denotes the probability that the system is in the initial state $i$ and $g_i(t)$ the probability that the system in this state
has not relaxed until time $t$.
For an uncoupled system and the presence of simple rate processes one has $g_i(t) := g_i^0(t) =  \exp(-\Gamma_i t)$. Since the
thermodynamics does not change upon coupling the values of the $p_i$ remain the same.
Furthermore, we define $G_i = \int dt \, g_i(t)$ and $\tau_\alpha = \sum_i p_i G_i$. The index 0
always denotes the uncoupled case. Note that $1/\langle \tau \rangle  = \langle \Gamma \rangle := \sum p_i \Gamma_i$.

For general $q$ one has to keep track of pairs of adjacent states (see above). However, a dramatic simplification is
possible in the limit of small $q$. Restricting oneself to the lowest order effect of the coupling one can systematically
neglect the forward-backward reactions which are at least of order $q^2$. Thus, one can employ a mean-field type approach
where a tagged system experiences fluctuations which are uncorrelated to the resulting passive exchange processes. These fluctuations
induce a reorganization of the tagged system with probability $q$ per relaxation process (more specifically: MB transition).
As a consequence the rate of change is $Q:= q/\langle \tau \rangle$. Formally, this can be written as \cite{Heuer_review}
\begin{equation}
(d/dt) g_i(t) = -\Gamma_i g_i(t) + Q (S(t)-g_i(t)).
\end{equation}
Whereas the first term on the right hand side describes the active relaxation of the tagged subsystem, starting in state $i$, the second term
characterizes a passive exchange process from some state $j$ ($j \ne i)$ to state $i$ due to the coupling to the adjacent subsystem.
Since we aim to keep the dependence on $Q$ only up to linear order, we can substitute $S(t)$ by $S^0(t)$. Then integration yields
\begin{equation}
-1 =  -(\Gamma_i + Q)G_i + Q \tau_\alpha^0.
\end{equation}
Thus, one can write
\begin{equation}
G_i = \frac{1}{\Gamma_i + Q}+ \frac{Q}{\Gamma_i + Q}\tau_\alpha^0 \approx \frac{1}{\Gamma_i}- Q \left (  \frac{1}{\Gamma_i^2} -\frac{\tau_\alpha^0}{\Gamma_i} \right ).
\end{equation}
Multiplication with $p_i$ and summation over all states yields
\begin{equation}
\tau_\alpha = \tau_\alpha^0 - Q \frac{(\tau_\alpha^0)^2}{\langle \tau \rangle } \left (  \frac{1}{(\tau_\alpha^0)^2}\langle 1/\Gamma^2 \rangle -1 \right ).
\end{equation}
Note that $\langle 1/\Gamma^2\rangle$ is identical to $\int dt \, t S^0(t)$. It is exactly the first term in the brackets which in \cite{Heuer_CTRW} has
been denoted as $1/\beta_M$ and is a measure for the non-exponentiality of the total relaxation, i.e. of the dynamic heteterogeneities. Thus, we can rewrite
\begin{equation}
\frac{\tau_\alpha}{\tau_\alpha^0} = 1 - q \frac{\tau_\alpha^0}{\langle \tau \rangle } \left ( \frac{1}{\beta_M} -1 \right ).
\end{equation}

More generally, in lowest order of $q$ the tagged system of $N_{min}$ particles will experience a coupling to the surrounding particles which is proportional to the number of particles,
i.e. $N - N_{min}$. Furthermore, after identification of $\tau_\alpha^0$ with $\tau_\alpha(N_{min})$  one may finally write for $N \ge N_{min}$ to lowest order in $q$
\begin{equation}
\frac{\tau_\alpha(N)}{\tau_\alpha(N_{min})} = 1 - q \left ( \frac{N}{N_{min}}-1 \right ) \frac{\tau_\alpha(N_{min})}{\langle \tau \rangle } \left ( \frac{1}{\beta_M} -1 \right ).
\end{equation}

Thus, after determination of $\beta_M$ and  $\tau_\alpha(N_{min})/\langle \tau \rangle$ this relation can be used to extract the dimensionless coupling constant $q$.

Note that this estimation breaks down if (i) the second term on the right hand side approaches one (then higher-order terms of $q$ matter) or (ii) $N$ starts to be much larger than $N_{min}$.
Then the coupling is no longer proportional to $N$ because the variation of distances between different subsystems starts to matter.

\end{document}